\documentclass[referee]{aa} 

\usepackage{graphicx}
\usepackage[utf8]{inputenc}
\usepackage[english]{babel}
\usepackage{natbib}
\usepackage{xcolor}
\usepackage{txfonts}
\begin{document}

   \title{Propagation of Waves above a Plage as Observed by IRIS and SDO}


   \author{P. Kayshap\fnmsep\thanks{Just to show the usage
          of the elements in the author field}
          \inst{1}
          \and
          A.K. Srivastava\inst{2}
          \and
          S.K. Tiwari\inst{3,4}
          \and
           P. Jel{\'i}nek\inst{1} 
           \and
           M. Mathioudakis\inst{5}
          }

   \institute{ University of South Bohemia, Faculty of Science, Institute of Physics, Brani\v sovsk\'a 1760, CZ -- 370 05 \v{C}esk\'e Bud\v{e}jovice, Czech Republic\\
              \email{virat.com@gmail.com}
              \and
            Department of Physics, Indian Institute of Technology (BHU), Varanasi, India
            \and
             Lockheed Martin Solar and Astrophysics Laboratory, 3251 Hanover Street, Building 252, Palo Alto, CA 94304, USA
            \and
            Bay Area Environmental Research Institute, NASA Research Park, Moffett Field, CA 94035, USA
            \and
            Astrophysics Research Centre, School of Mathematics and Physics, Queen's Univeristy, Belfast, BT7 1NN, UK.
             }

   \date{}
 
  \abstract
   {MHD waves are proposed to transport sufficient energy from the photosphere to heat the transition-region (TR) and corona. However, various aspects of these waves  such as their nature, propagation characteristics and role in the atmospheric heating process remain poorly understood and are a matter of further investigation.}
   {We aim to investigate wave propagation within an active-region (AR) plage using IRIS and AIA observations. The main motivation is to understand the relationship between photospheric and TR oscillations. We plan to identify the locations in the plage region where magnetic flux tubes are essentially vertical, and further our understanding of the propagation and nature of these waves.}
{ We have used photospheric observations from AIA (i.e., AIA 1700~{\AA}) as well as TR imaging observations (IRIS/SJI Si~{\sc iv} 1400.0~{\AA}). We have investigated propagation of the waves into the TR from the photosphere using wavelet analysis (e.g., cross power, coherence and phase difference) with inclusion of a customized noise model.
   }
   {Fast Fourier Transform(FFT) shows the distribution of wave power at photospheric $\&$ TR heights. Waves with periods between 2.0- and 9.0-minutes appear to be correlated between the photosphere and TR. We exploited a customized noise model to estimate 95\% confidence levels for IRIS observations. On the basis of the sound speed in the TR and estimated propagation speed, these waves are best interpreted as the slow magneto acoustic waves (SMAW). It is found that almost all locations show correlation/propagation of waves over broad range of period from photosphere to TR. It suggests the wave's correlation/propagation spatial occurrence frequency is very high within the plage area.}
   {}


   \maketitle
%

\section{Introduction}
Understanding wave propagation is a very important topic in solar physics as these waves can transport energy into the upper layers of the Sun's atmosphere. The energies carried by these waves can play a  crucial role in the heating of the interface region and inner corona. The interface-region heating can not be fully understood without understanding/characterizing the wave propagation and the effect of the complex solar atmosphere on it. The plasma conditions, perturbations, and structured magnetic fields, lead to a complex behavior of waves in magnetic flux-tube. Using observations from different instruments as well as numerical simulations, there are several reports that shed light on various aspects of the waves (e.g., origin, properties, and dynamics) and their propagation throughout the solar atmosphere (\citealt{DeMoortel2000,DeMoortel2002,DePon2003,Centeno2006,Centeno2009,Jel2009,Jess2012,Jel2013,Jel2015,Prasad2015,Murawski2018,Kayshap2018}). In sunspot umbra, 3-minute waves propagate up to the chromosphere from the photosphere as reported by \cite{Centeno2006, Centeno2009} using spectropolarimetric observations. Interface-Region Imaging Spectrometer (IRIS) observations have shown that the  3-minute oscillations can propagate within umbra up to the TR and corona (\cite{Tian2014, Kho2015}). In magnetic "free-regions" (i.e., inter-network), the propagation of 3-minute waves is widely reported (e.g.,\citealt{Lites1982,Wik2000,Judge2001,Bloomfield2004,Kayshap2018}). Using IRIS high-resolution observations, \cite{Kayshap2018} explored the propagation of 3 minute oscillations in the inter-network.  The propagation of low frequencies (5-minute) from the photosphere to the higher layers have also been reported in a small magnetic patch by \cite{Srivastava2008}.\\
In coronal loops, \cite{DeMoortel2002} have proposed that the photospheric oscillations (specifically, 3 $\&$ 5 minutes) can reach up to the TR/coronal heights.  
It has been proposed that these are different manifestations of slow magneto acoustic waves (SMAW) in coronal loops (\citealt{Jess2012,Prasad2015}). Using TRACE observations, it is has also been reported that photospheric power (p-mode) can reach the TR  within the plage atmosphere (\citealt{DePon2005}) and the interaction of p-mode with the magnetic field generates magnetohydrodynamics (MHD) wave modes (\citealt{SP1991,Cally1994,Jess2015}). The magnetic field acts as a guide for the waves to reach up to coronal heights (\citealt{Cally2007, DePon2003,DePon2005}).\\
Almost all the works that investigate wave propagation use wavelet analysis for the study of phase relations. The first and most fundamental necessity is has to do with the reliability of the wave periods that are detected through wavelet analysis. The reliability of the period depends on the confidence levels that are measured with the help of the assumed theoretical spectrum. The white noise (i.e., flat spectrum with no frequency dependency) is most widely used theoretical spectrum to calculate the confidence levels. However, the white noise does not represent the true noise inherited in the signal of the solar atmospheric. The use of an incorrect noise model may lead to the false detection of wave periods (\citealt{Auch2016, Thr2017}). It has been reported that the Fourier spectra of coronal signals behave as a power-law  (e.g., \citealt{Auch2014, Gupta2014, Inglis2015, Ireland2015}). The power-law like nature of Fourier/wavelet spectra (i.e., P $\propto$ f$^{-\alpha}$) is characteristic of red noise when the exponent value is -2. The exponent values may vary from one time-series to another. Therefore, we emphasize the importance and necessity of the power-law model for the estimation of the confidence levels (e.g., 
\citealt{Gab2002,Vau2005,Auch2016,Pugh2017}). Finally, we would like to mention that \cite{Auch2016} have beautifully justified the necessity of power law (along with kappa function and constant background) to detect wave periods in coronal loops. Our work follows \cite{Auch2016} to calculate the confidence levels using a power-law function. \\
The main motivation of our work is to identify the nature of wave motions and their propagation properties in an active-region plage area using high-resolution observations from IRIS (\citealt{DePon2014}). The work is organized as follows. Sect.~2 describes the observational data and analysis. Sect.~3 describes the deduced results under three different subsections. The first subsection (Sect.~3.1) is dedicated to Fourier maps of plages in the different frequency bins. The second subsection (Sect.~3.2) describes the wavelet analysis and associated results using SDO/AIA 1700~{\AA} (\citealt{Lemen2012}) and IRIS/SJI 1400~{\AA}. In the final subsection (Sect.~3.3), we investigate the nature of SMAW waves in the plage region. In the last section, discussion and conclusions are outlined. In addition, appendix-A discusses the diagnostics that dominate the emission in the IRIS/SJI 1400~{\AA} channel.
\section{Observational Data $\&$ Analysis}
IRIS observed an active-region plage on 28 July 2014 from approximately 17:59 UT to 19:52~UT. IRIS captures the solar spectra in the near and far ultraviolet, which includes many photospheric and chromospheric/TR lines, e.g., Mn~{\sc i} 2803.8~{\AA}, Mg~{\sc ii} k 2796~{\AA}, C~{\sc ii} 1334.53~{\AA}, Si~{\sc iv} 1393.75~{\AA}. 
In this study we have used mainly the IRIS/Slit-Jaw Imager (SJI) at 1400~{\AA}, Si~{\sc iv} 1393.75~{\AA}, and AIA 1700 ~{\AA}. In addition, we have also utilized magnetic field inclination and line-of-sight (LOS) information from HMI. At the time of the observations, the AR plage was located at disk center (i.e., very close to $\mu$ = 1.0). The on-disk observations minimize projection effects (\cite{Falconer2016}). However, another possibility of projection effects comes due to the inclination of magnetic field which can lead to offsets at different heights.\\
AIA 1700~{\AA} imaging observations sample the emission from the upper photosphere, while IRIS/SJI Si~{\sc iv} 1400~{\AA} is dominated by a spatially varying mix of upper photospheric continuum and low TR emission. The IRIS/SJI 1400~{\AA} filter is a broadband filter with a width of 55~{\AA}, and is dominated by continuum emission in the vicinity of two strong Si~{\sc iv} lines. Therefore, this filter can also capture some photospheric emission. In order to check the parts of the atmosphere that contribute to the emission in the IRIS/SJI (i.e., photosphere or TR), we have investigated the IRIS/SJI 1400~{\AA} observations in conjunction with Si~{\sc iv}. Using this analysis we found that the IRIS/SJI 1400 filter is dominated by the TR emission in the plage area. However, we did not find a well defined correlation between SJI 1400 and Si~{\sc iv} line in the surrounding quiet-Sun. This suggests that our region of interest (ROI) is dominated by the TR emission. More information is provided in appendix A\\
   \begin{figure*}
   \centering
   \includegraphics[trim = 2.0cm 3.2cm 2.0cm 2.0cm,scale=1.3]{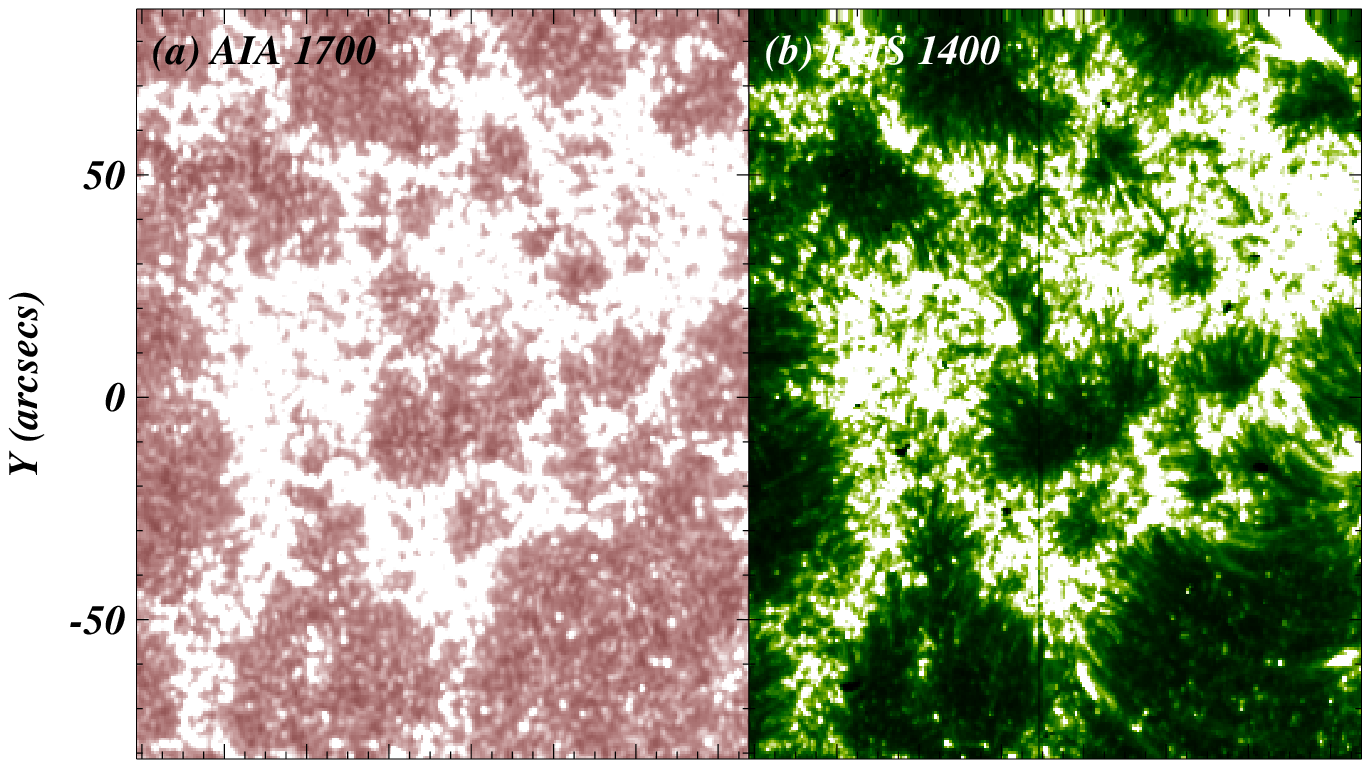}
   \includegraphics[trim = 2.0cm 1.0cm 2.0cm 2.0cm,scale=1.3]{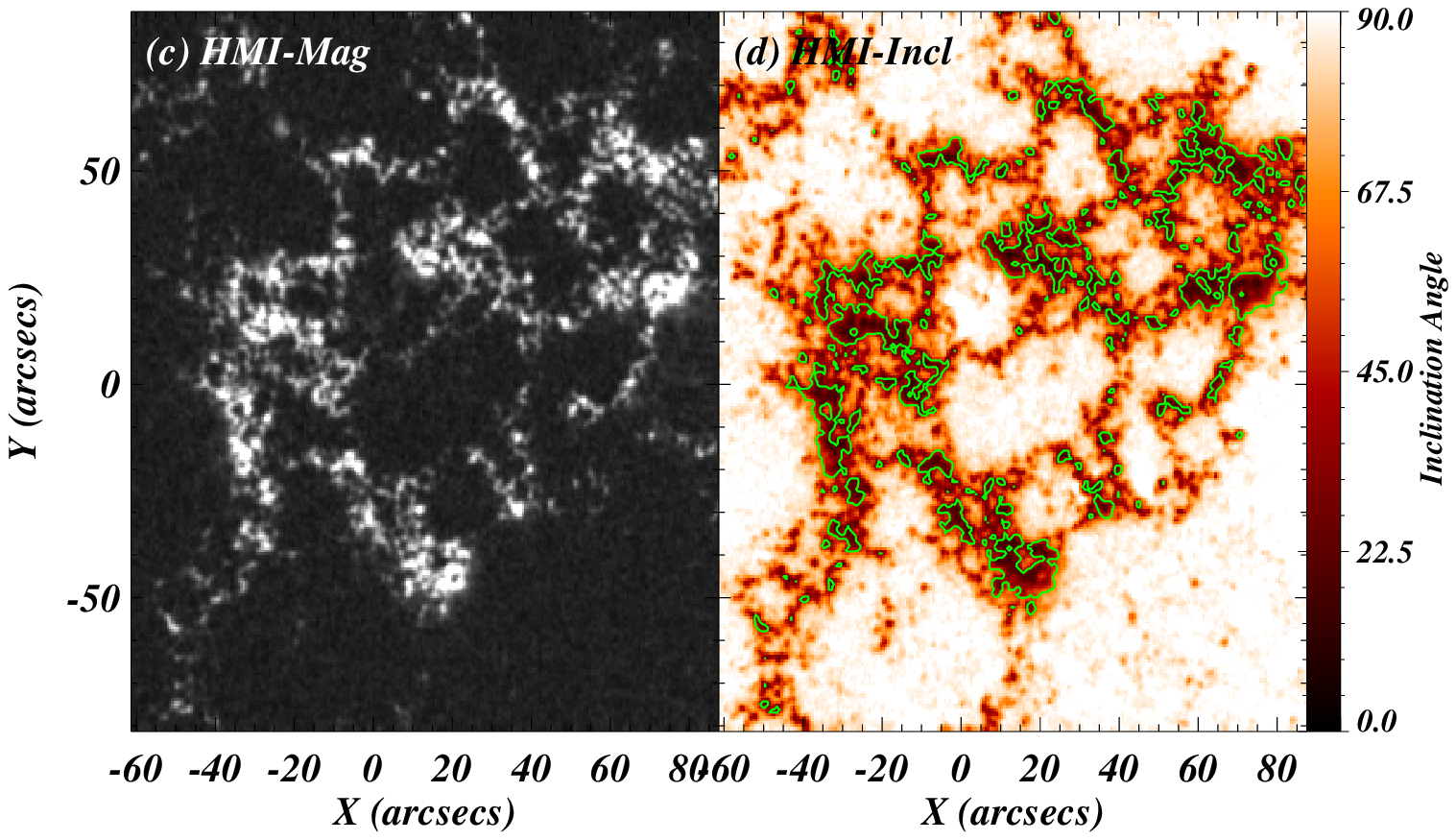}
   \caption{The photospheric (SDO/AIA 1700~{\AA}; panel a) and TR (IRIS/SJI 1400~{\AA}; panel b) images of the region-of-interest (plage region). The plage is seen as high-intensity patches clearly visible in the photosphere as well as TR (panel a $\&$ b). We show the absolute line-of-sight (LOS) magnetic field (panel c) and magnetic field inclination map (panel d) of the same region using SDO/HMI observations. High magnetic field and low inclination angles are visible within the vicinity of the plage area. We have selected the plage locations using a threshold of absolute magnetic field (i.e., above 300 Gauss) and locations are contoured (i.e, above 300 Gauss) over magnetic filed inclination (panel d) by green color which reveal the presence of almost vertical (or very low inclination) fields.}
    \label{fig:ref_image}%
    \end{figure*}
Figure~\ref{fig:ref_image} shows AIA 1700~{\AA} (panel a) and IRIS/SJI 1400~{\AA} (panel b) images of the observed plage region.  We also show the LOS photospheric magnetic field (panel c) and field inclination angles (panel d). Strong magnetic fields are evident in the vicinity of the plage. The inclination angle from these plage locations varies from zero (i.e., vertical magnetic field in the central areas) to 70-80 degrees (at the outer edges). However, we are interested in the vertical magnetic field locations within these plage areas. Therefore, we have put a threshold of magnetic field (i.e., higher than 300 G) to identify the vertical magnetic field plage locations (VMPLs). The identified plage locations are overdrawn by green contours on the inclination map (panel d). The inclination angles are very low within the VMPLs, which justifies that the magnetic field is essentially vertical. After identification of VMPLs, we extracted the time-series (TS) from AIA 1700~{\AA} and IRIS/SJI 1400~{\AA} data.
\section {Results}
We investigate wave propagation between the photosphere and the TR in the VMPLs. We utilize wavelet transform (e.g., cross wavelet transform, coherence, and phase difference) for this work. However, before presenting the wavelet analysis $\&$ the associated  results, we present Fourier power maps of the plage region. This provides the wave power distribution within the plage-region at two different heights (i.e., AIA~1700~{\AA} and IRIS~1400~{\AA}).
\subsection{Plage's Fourier Maps} Fast Fourier Transform (FFT) of the plage observations is performed in AIA~1700~{\AA} (i.e., photosphere) and IRIS~1400~{\AA} (TR). The FFT provides the power spectral density (i.e., wave power vs frequency) over a particular solar region, e.g., plage in the present work.
  \begin{figure*}
   \mbox{
   \includegraphics[trim = 0.0cm 2.0cm 0.0cm 1.5cm,scale=0.85]{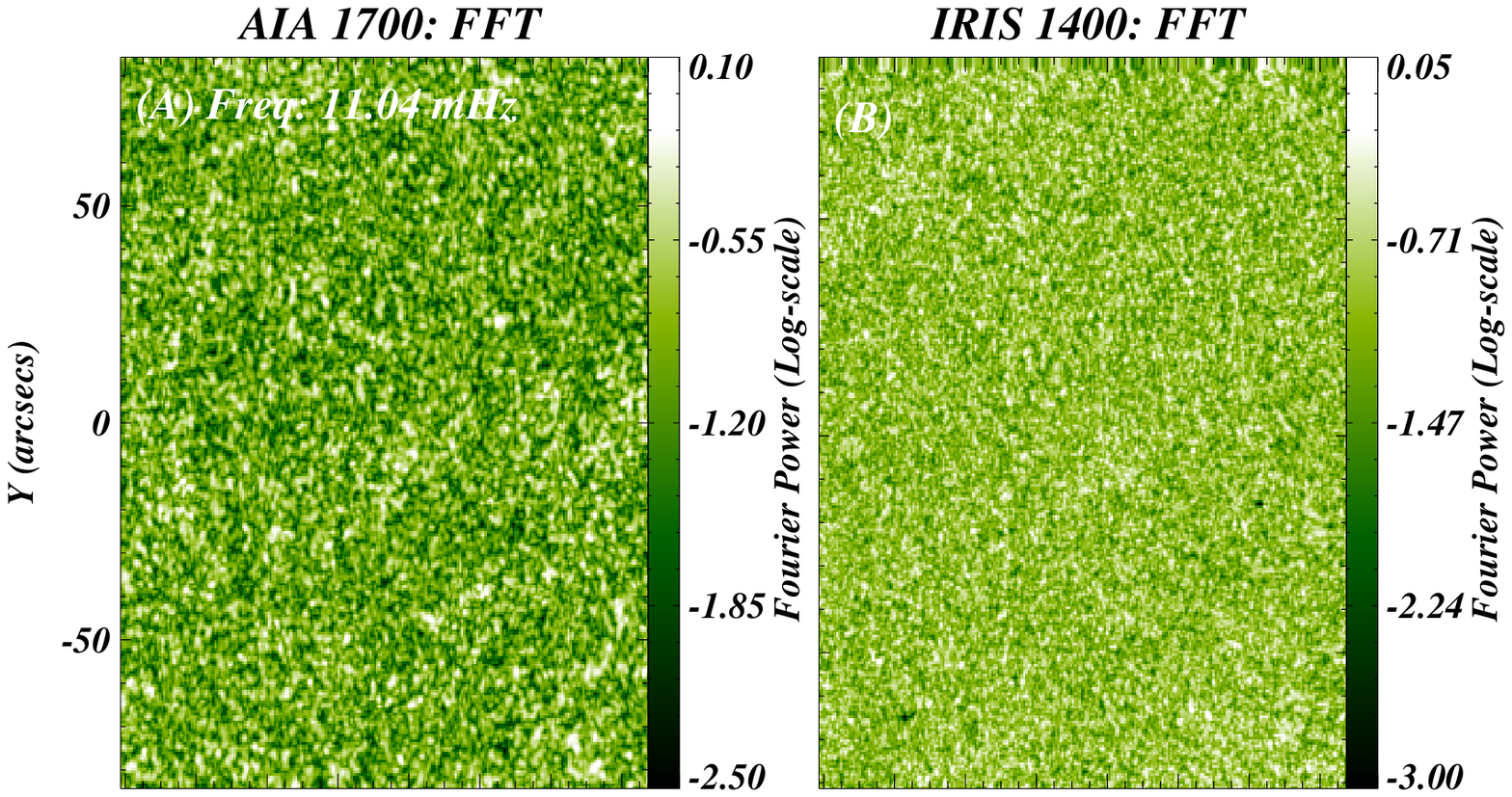}
   }
 \mbox{
   \includegraphics[trim = 0.0cm 2.0cm 0.0cm 2.0cm,scale=0.85]{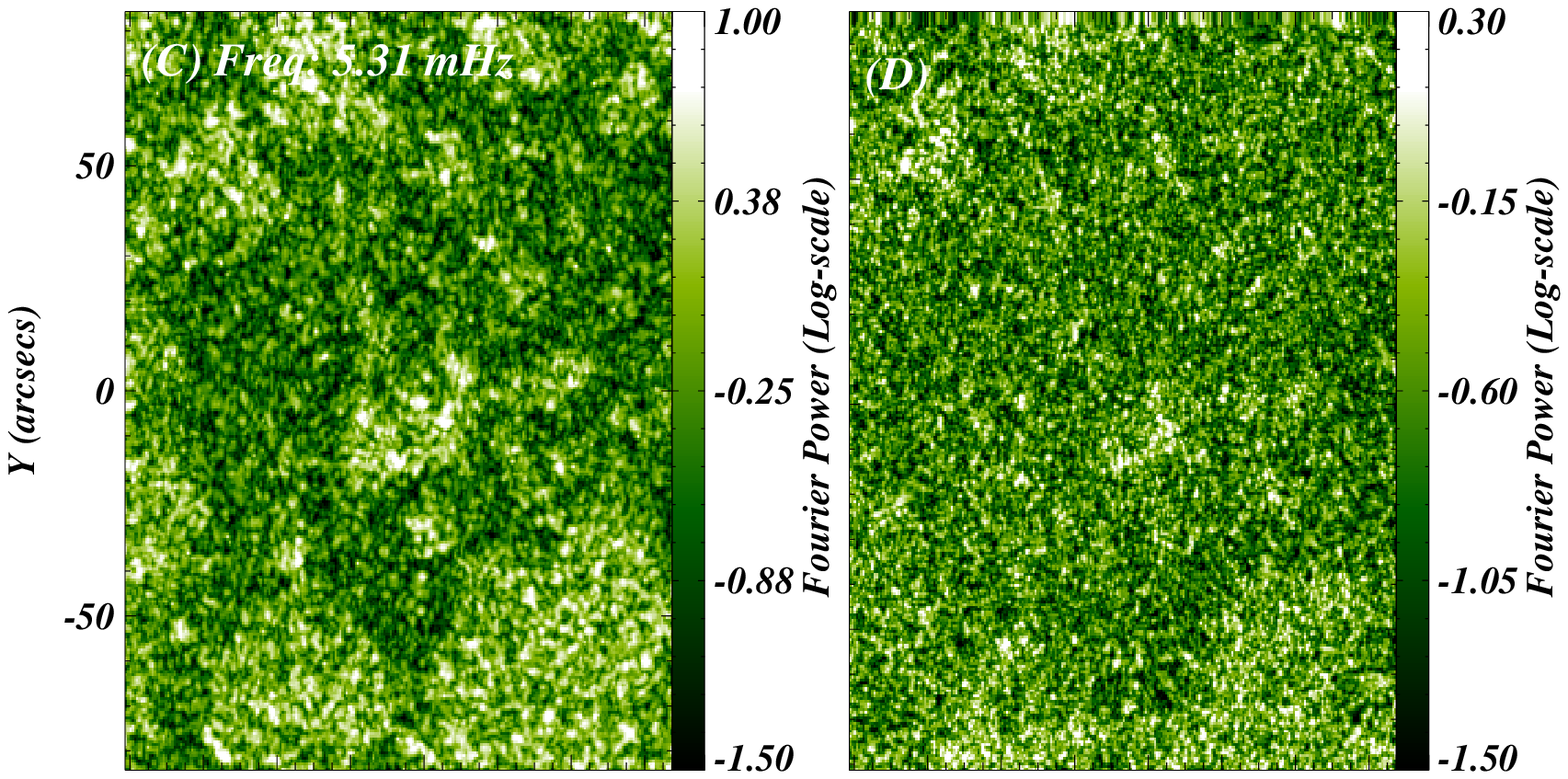}
}
\mbox{
   \includegraphics[trim = 0.0cm 1.0cm 0.0cm 2.0cm,scale=0.85]{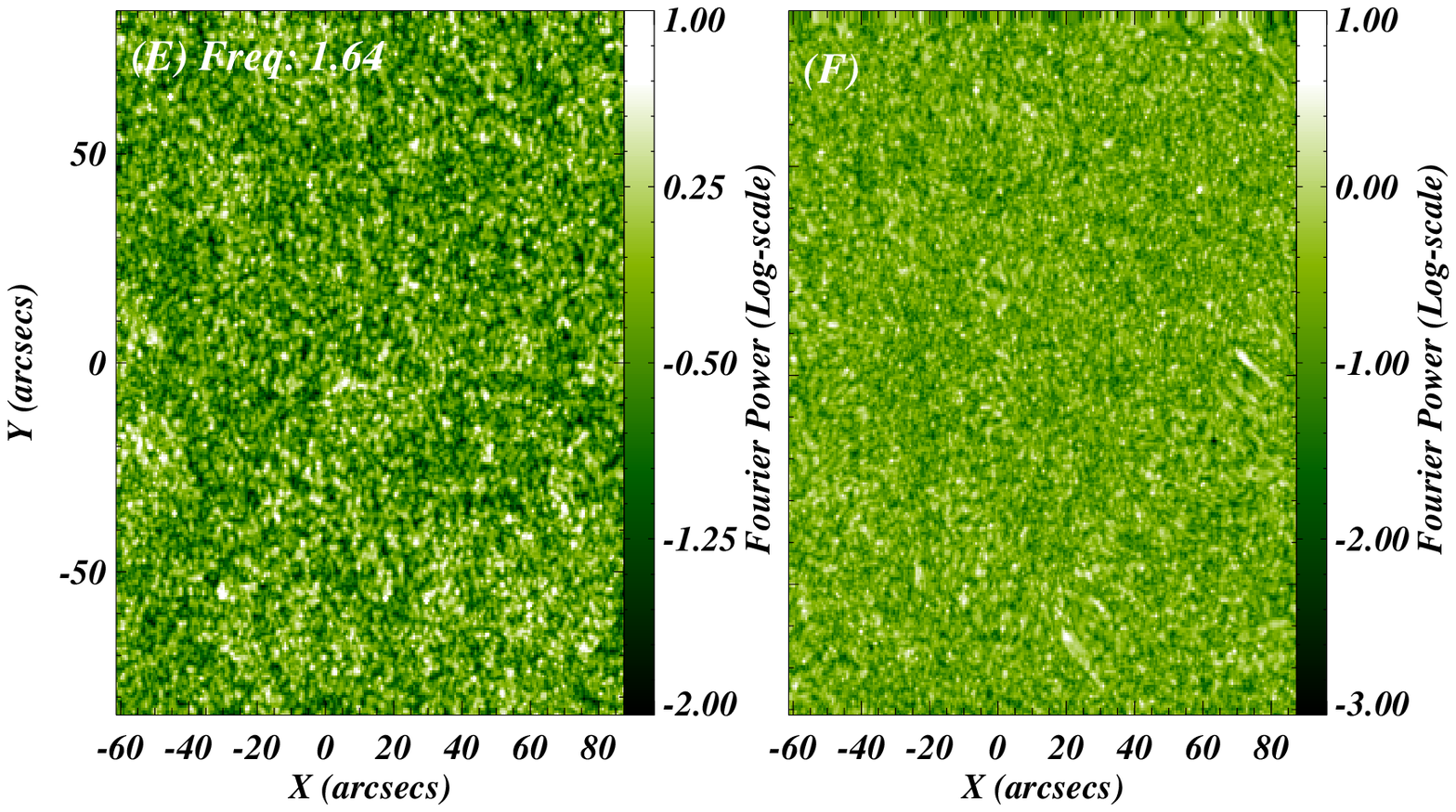}
   }
   
   \caption{Fourier power maps (log-scale) within the different frequency ranges are shown for AIA~1700~{\AA} (left column) and IRIS~1400~{\AA} (right column). Panel A and B show the Fourier power map in high frequency range (i.e., 11.04 mHz) for the photosphere and TR. While, the middle row panels displays Fourier power map in intermediate frequency (i.e., 5.37 mHz) for photosphere (panel C) and TR (panel D). The bottom row shows Fourier maps but for low frequency (i.e., 1.64 mHz).}
    \label{fig:fourier_map}%
    \end{figure*}
We have normalized the Fourier power as explained in the way \cite{Gabriel2002} and used by \cite{Froment2015} and \cite{Auch2016}. Based on this methodology we estimate the scaling factor (i.e., m as defined in previous works) corresponding to the used confidence levels. We estimate the scaling factor value to be 8.41 for confidence levels of 95\%, \textbf{i.e., if the power is higher than 8.41 then the confidence level is above 95\%}. Using the value of the scaling factor (i.e., m) and estimated noise we derived the final array (i.e., m*estimated noise) to normalize the Fourier power for each pixel. Finally, we estimated the normalized power maps for each frequency at each location in the observed region. In order to highlight the wave power distribution in the observed region, we selected three different frequency ranges, namely,(1) high-frequency range  {--} 11.04 mHz (period: 1.5 minutes), (2) intermediate frequency range {--} 5.31 mHz (period: 3.10 minutes), and (3) low frequency range {--} 1.64 mHz (periods: 10.15 minutes). We extracted 2-D Fourier maps (for these 3 selected frequencies) for AIA~1700~{\AA} and IRIS~1400~{\AA}. The 2-D Fourier maps for both channels are shown in Fig.~\ref{fig:fourier_map}.\\
\textit{High Frequency:} In Fig.~\ref{fig:fourier_map}, we have shown the Fourier power map in the high frequency range (i.e, 11.04 mHz) for AIA~1700~{\AA} (panel a) and IRIS~1400~{\AA} (panel b). The observed area contains mostly a plage region along with the surrounding region at the edges of images (see Fig.~\ref{fig:ref_image}). The normalized Fourier power maps show uniform distribution of the power of the waves in the plage and the surrounding area at both heights in the solar atmosphere (e.g., AIA 1700~{\AA} $\&$ IRIS 1400~{\AA}). Therefore, we do observe that the behavior of the plage $\&$ surrounding region is almost similar at this high-frequency. However, it should be noted that power is very low at the photosphere (panel A) and TR (panel B). Furthermore, TR power is slightly lower than the photospheric power.\\
\textit{Intermediate Frequency:} The middle row of Fig.~\ref{fig:fourier_map} shows Fourier power map of the photosphere (panel C) and TR (panel D) for intermediate frequency (i.e., 5.31 mHz). Particularly, this is in the frequency regime where most of the wave power is inherited and propagates into the higher layers. The pattern of photospheric wave power at intermediate frequency is completely different from the behavior that we found in high frequencies. We see that power in the surrounding region is marginally higher than the inherited power in the plage. A similar behavior exists in TR heights, however, the pattern is not as prominent as at the photospheric level.\\
\textit{Low Frequency:} The bottom row of Fig.~\ref{fig:fourier_map} shows the Fourier power map of the photosphere (panel E) and TR (panel F) at low frequency (i.e., 1.64 mHz). At this frequency range, we do not see any difference between the plage and surroundings at the photospheric (panel E) and TR levels (panel F).   
\subsection{Wavelet Analysis} A wavelet transform involves the convolution between the TS and the "mother" function. There are different type of "mother" functions with the Wavelet transform (e.g., Morlet, Paul and Derivative of Gaussian (DOG)). For the present analysis, we have used the Morlet mother function with a dimensionless frequency $\omega_{0}$ = 6, which is suitable for investigating the propagation of waves with a range of frequencies. The Morlet function consists of the plane wave which is modulated by a Gaussian window. 
  \begin{figure*}
   \mbox{
   \includegraphics[trim = 2.5cm 0.0cm 2.0cm 0.0cm,scale=0.8]{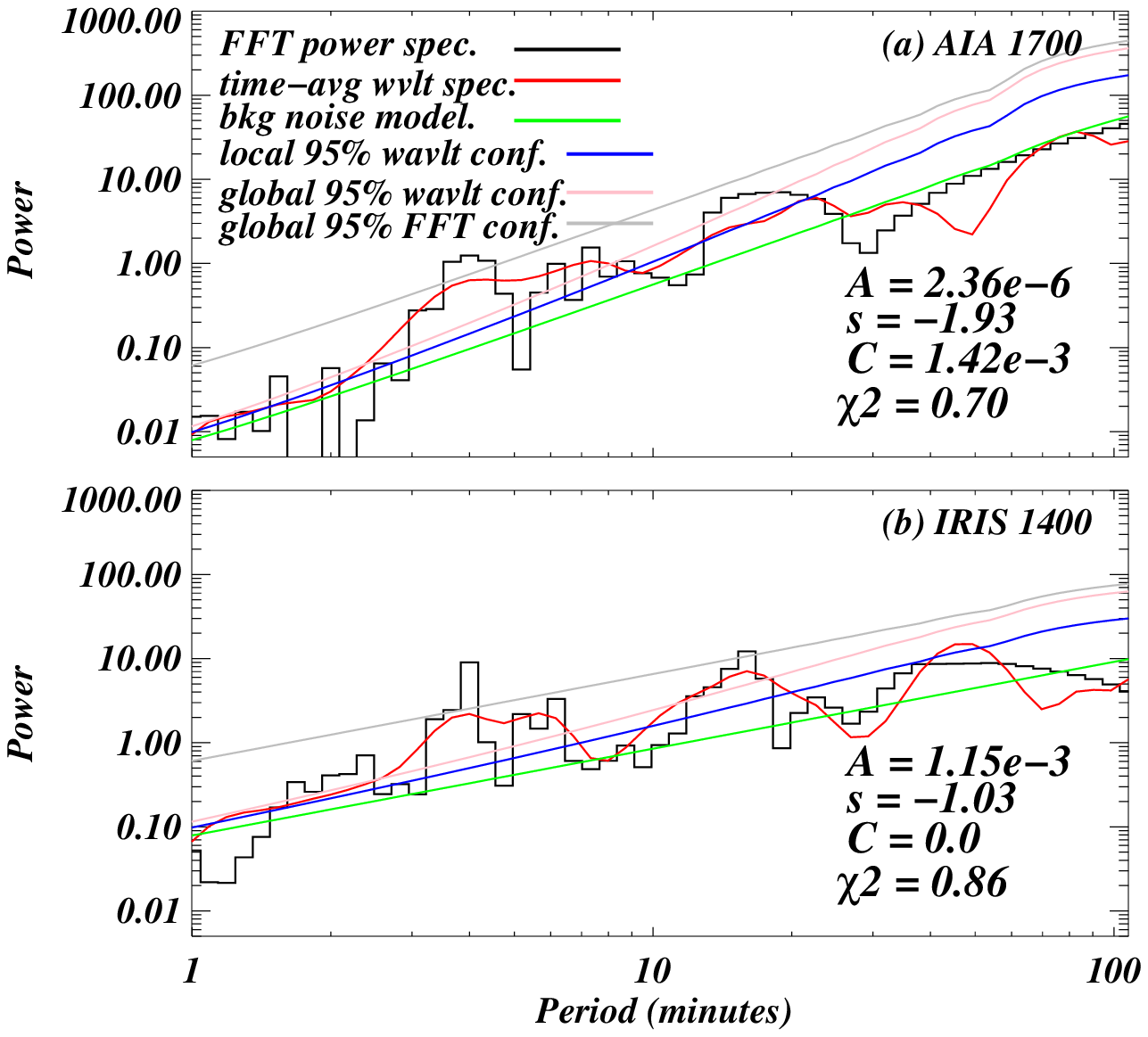}
   \includegraphics[trim = 4.5cm 0.0cm 2.0cm 0.0cm,scale=0.8]{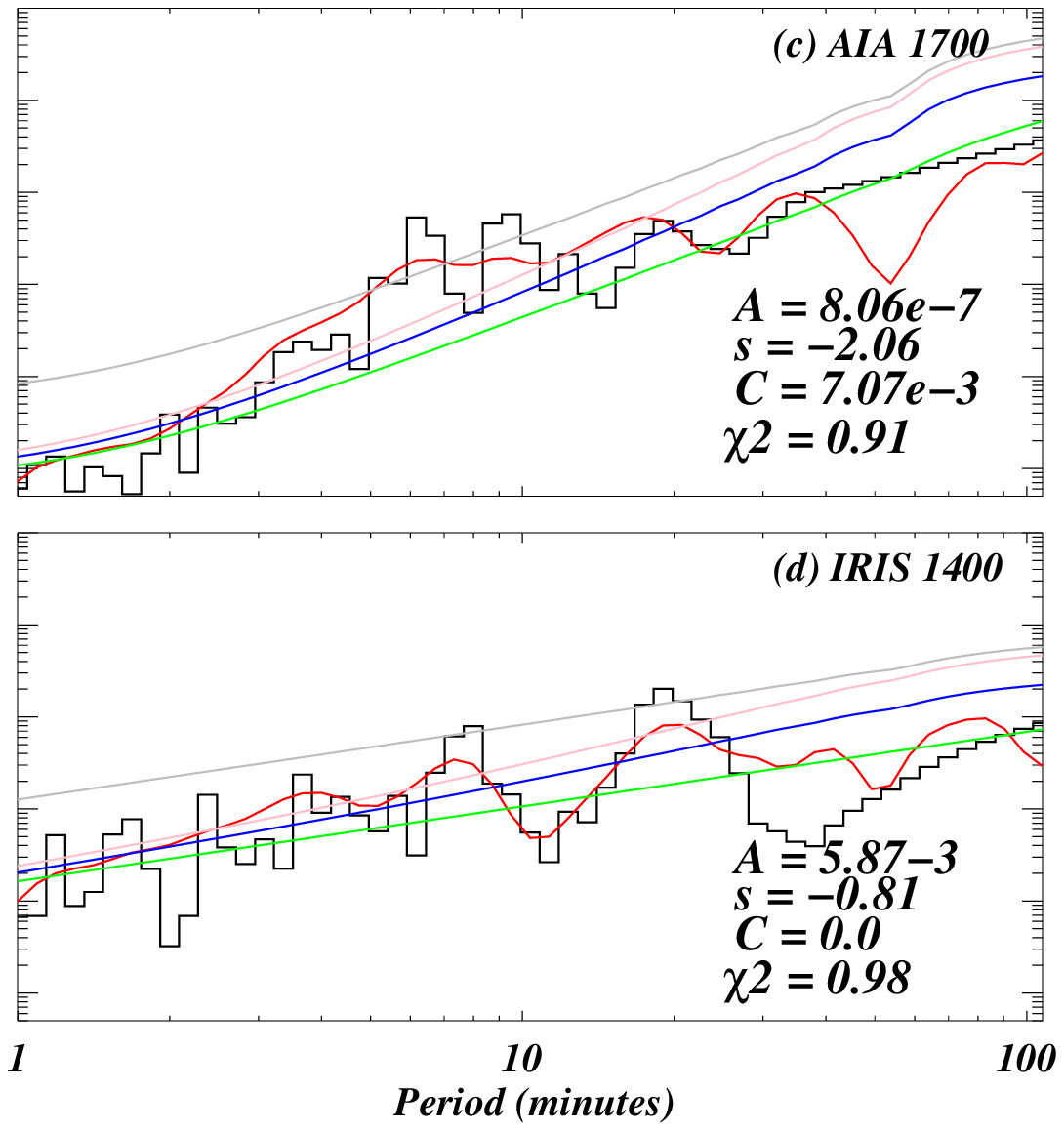}
   }
   
   \caption{The generic noise model for AIA 1700 (panel a and c) and IRIS/SJI 1400 (panel b and d). The FFT power spectrum (black line), time averaged wavelet spectrum (red line), and fitted noise model (green line) are shown in each panel. Using fitted noise, we have drawn local and global 95\% confidence levels, which are shown by blue and pink colors respectively. The parameters used for the fits parameters also provided. The low $\chi^2$ values justify the reliability of the fitted noise model on the FFT power spectrum.}
    \label{fig:noise}%
    \end{figure*}
  
\begin{equation}
\label{eq:morlet}
{{\psi_{0}(\eta)}} = \pi^{-1/4} e^{i\omega_{0}\eta} e^{-\eta^2/2}\, 
\end{equation}     
Eq.~\ref{eq:morlet} describes the Morlet function,where $\omega_{0}$ and $\eta$ are the non dimensional frequency and time parameters. The wavelet transform provides a 2-D complex array for a TS. The power of the wavelet is defined as the square of absolute magnitude of the complex wavelet.
  \begin{figure*}
   \centering
    \mbox{
   \includegraphics[trim = 3.0cm 0.0cm 2.0cm 0.1cm,scale=0.9]{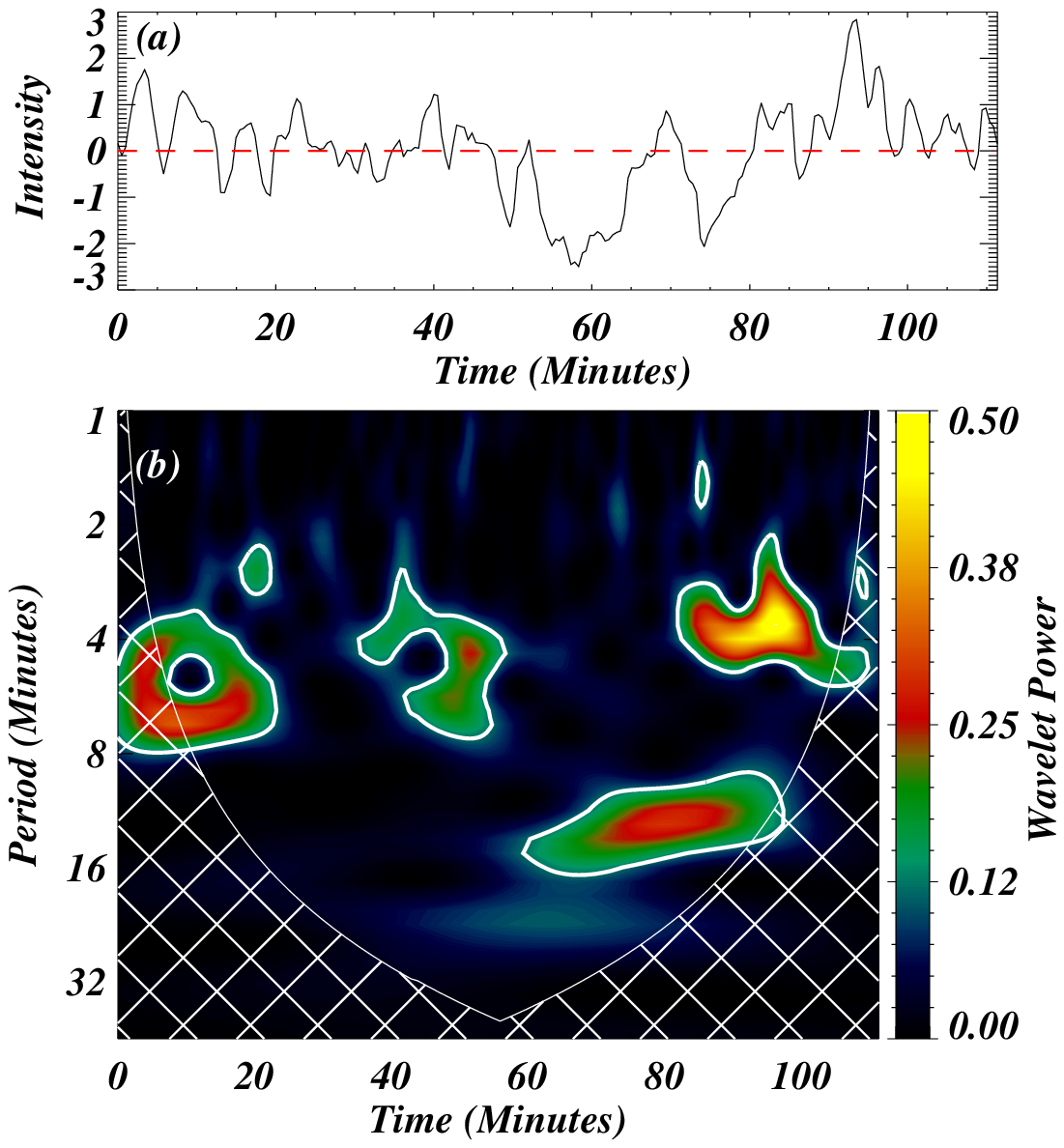}
   \includegraphics[trim = 4.7cm 0.0cm 2.0cm 0.1cm,scale=0.9]{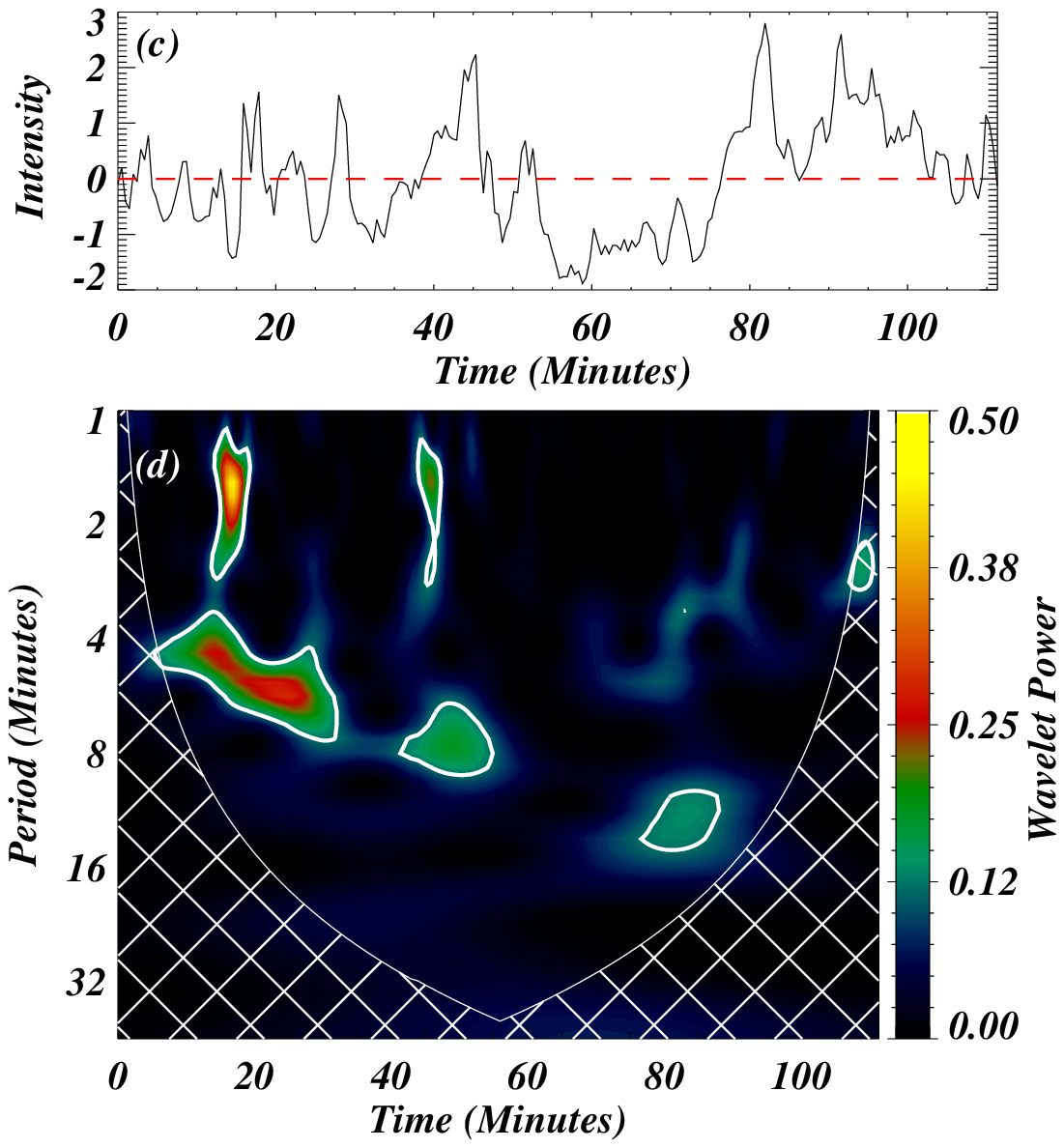}
   }
   \caption{Panel (a) shows the AIA 1700~{\AA} TS while panel (b) displays the corresponding normalized wavelet power. The photospheric power is mainly concentrated in the broad range (i.e., 2-9 minutes) as outlined by white contours (i.e, 95\% global confidence levels). Similarly, panel (c) and (d) show the IRIS/SJI 1400~{\AA} TS and its corresponding normalized wavelet power, respectively. The power within the TR is distributed within the period of 2-8 minutes. The white contours on each wavelet map (panel b $\&$ d) outlines the 95\% global confidence levels. The cross-hatched white area outlines the COI.}
    \label{fig:wavelet}    
    \end{figure*}
We have shown intensity TS and wavelet power maps (i.e., |W(s)|$^2$) for the photospheric (panel a) and TR (panel c; Fig~\ref{fig:wavelet}) emission from a location within the selected plage locations. We have not applied any  smoothing before the wavelet transform as it can add spurious periodicities in the TS \citep{Auch2016}. We have normalized the wavelet power as described in subsection 3.1. A major fraction of the photospheric power is concentrated in periods shorter than 15 minutes (i.e., high frequency), however, significant power can also be found around 5 minutes and reflects the well know photospheric oscillations (see panel; Fig.~\ref{fig:wavelet}). In the TR we also see the dominant power in high periods  (panel d; Figure~\ref{fig:wavelet}). This  power is not spurious and is persistent over long timescale (i.e., such periods exist more than 25 minutes or even longer, cf., Fig~\ref{fig:wavelet}). It indicates wave propagation from the photosphere into the TR. The white hatched area represents the Cone-of-Influence (COI) on each wavelet power map (cf.,Fig.~\ref{fig:wavelet}) while the solid white line outlines the 95\% confidence levels.\\ 
\begin{figure*}
   \mbox{
   \includegraphics[trim = 2.0cm 0.0cm 2.0cm 0.0cm,scale=0.62]{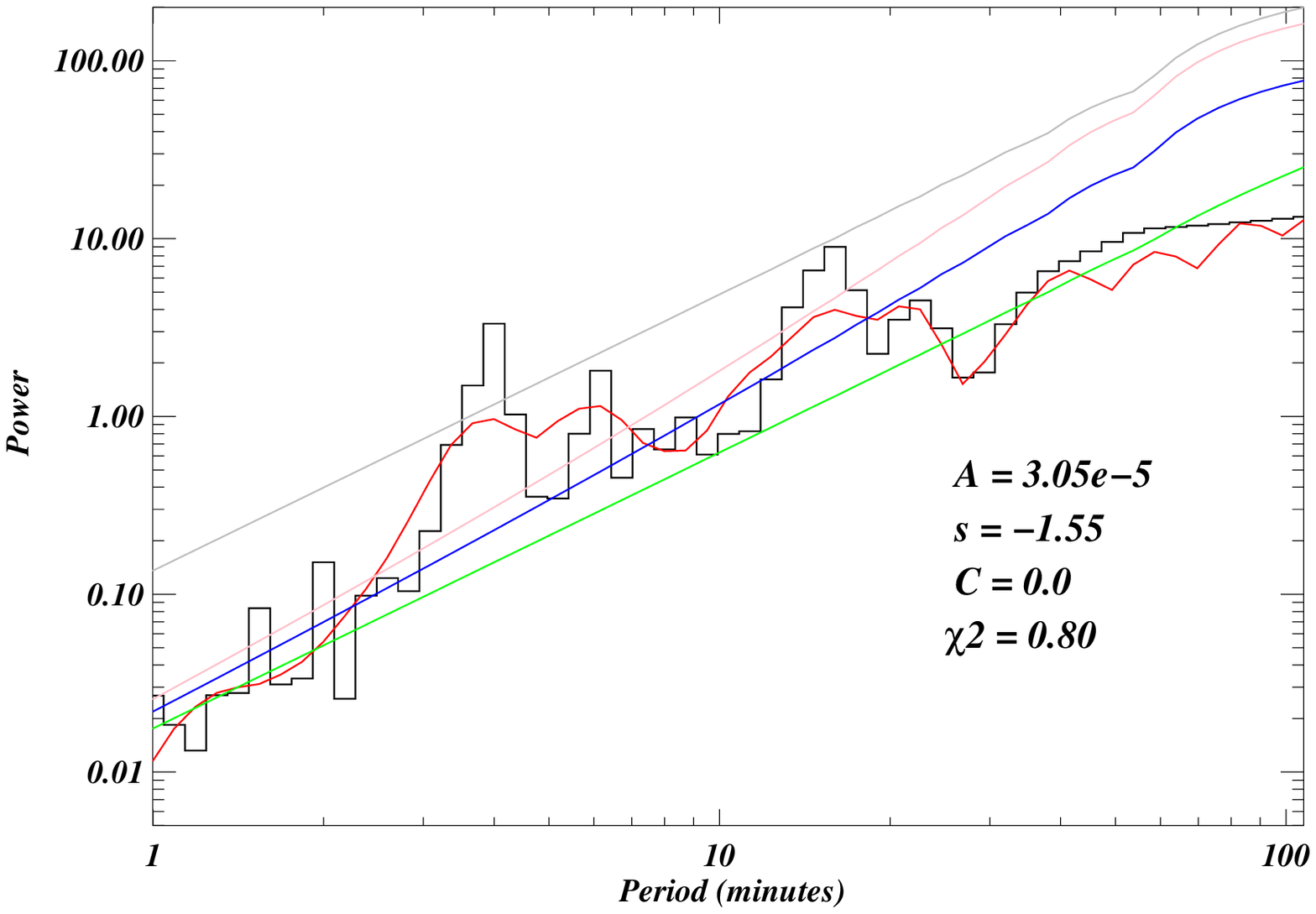}
   \includegraphics[trim = 0.5cm 0.0cm 2.0cm 0.0cm,scale=0.62]{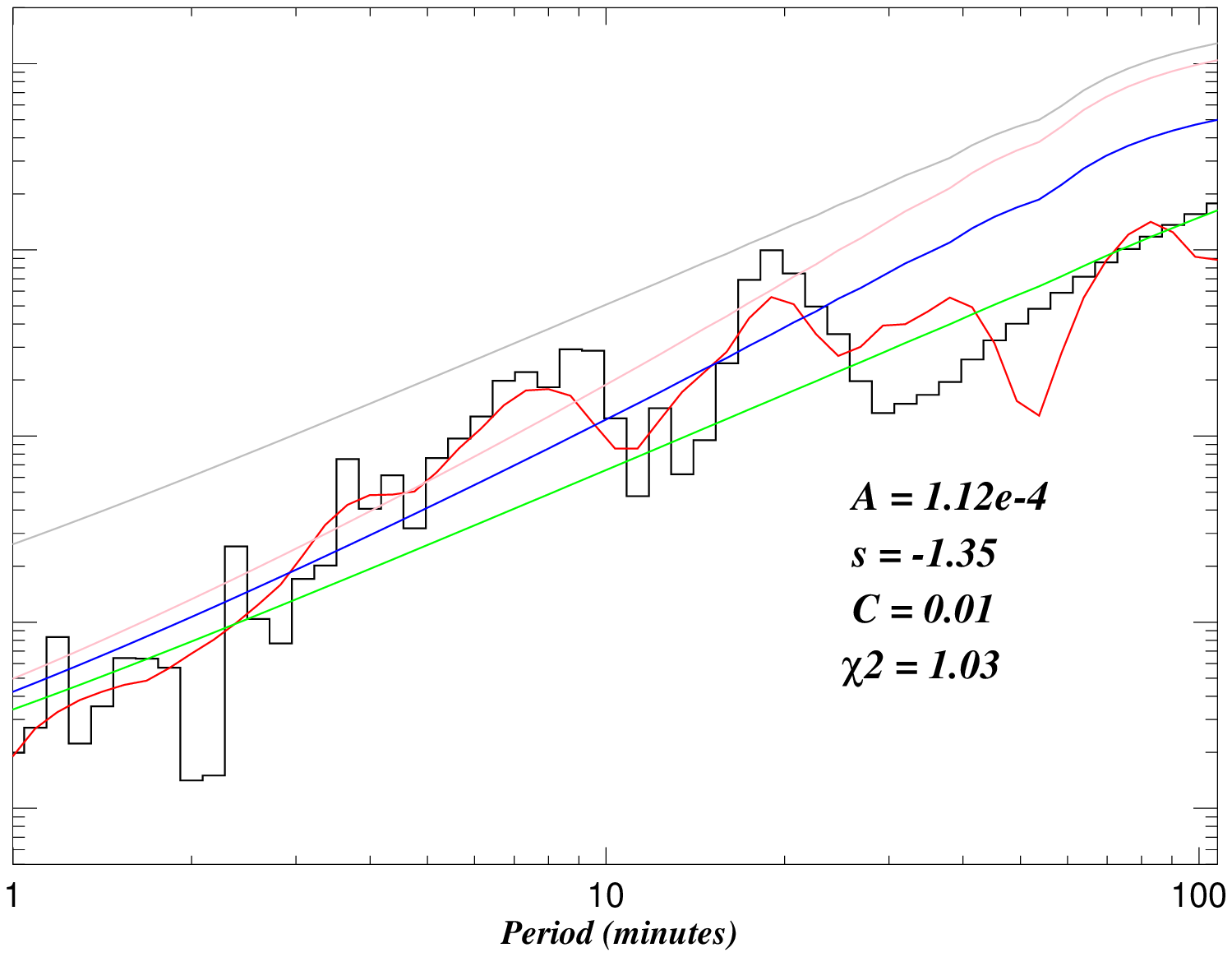}
   }
\caption{Similar to Fig.~\ref{fig:noise} but for cross spectra drawn using IRIS/SJI 1400~{\AA} and AIA 1700~{\AA}.}
\label{fig:cross_noise}%
\end{figure*}
The red or white noise spectrum is often used to calculate the confidence levels as described by \cite{Torr1998}. However, it should be noted that red or white noise does not really capture the inherited noise within the TS, which can in turn mislead the estimation of the confidence levels \citep{Auch2016}. Ultraviolet (UV) and extreme ultraviolet (EUV) light curves tend to exhibit an overall power-law behavior and the inbuilt red noise model is inappropriate to fit the spectrum (except if the power-law exponent is -2.0; private communication with De Pontieu). In the present analysis, we have also found that the TS shows a power-law behavior with a broad distribution the exponent values and specific value of the power-law exponent (i.e., -2) are rare. Therefore, we can say that in-built red-noise  model is not appropriate to estimate the confidence levels. 
Recently, \cite{Auch2016} proposed a generic noise model for an EUV TS (using AIA high-temperature filter observations) and have also shown that smoothing can add artificial periodicities. After \cite{Auch2016}, \cite{Thr2017} have also adopted the same generic noise model for EUV TS under different physical conditions of solar atmosphere. \cite{Auch2016} proposed that the generic noise model is a linear combination of the power law, kappa function, and white noise. It is shown that the coronal signals are well fitted through this model. The present work deals with the signal from the lower solar atmosphere (i.e., photosphere $\&$ TR), therefore, the noise characteristics may be different from the inherited noise in the coronal time-series. We check this fact and found that signals used in this work are best represented by only by a power law. Hence, we utilized only a power-law function and omitted the kappa function $\&$ white noise contribution. We have fitted a function $\sigma$(v) to each power spectrum which (i.e., $\sigma$(v)) is described as follows
\begin{equation}
\label{eq:noise_model}
{\sigma(v)}= Av^s+C
\end{equation} 
Although the contribution from constant (i.e., C {--} white noise) is low in most of the conditions still it may play an important role for some cases. Therefore, we kept the constant term in our model.
We first fitted $\sigma$(v) of each photospheric and TR power spectrum. Two samples are shown in the Figure~\ref{fig:noise} with various components. Each panel of Figure~\ref{fig:noise} shows the FFT power spectrum (black histogram), time averaged wavelet spectrum (red line) and the fitted generic noise model ($sigma$(v); green line). Using this noise model, we estimate the local (blue line in both panels of Fig~\ref{fig:noise}) and global 95\% Fourier confidence level (pink line in both panels of Fig.~\ref{fig:noise}). We also estimated the global 95\% confidence levels which are shown by the gray solid line in both panels of Fig.~\ref{fig:noise}. It should be noted that the global 95\% Fourier confidence level has high values compared to the local/global 95\% wavelet confidence level for all periods. 
The local/global 95\% confidence levels lie below the power spectra (i.e, Fourier $\&$ time-averaged wavelet) for most of the period bins. However, the global 95\% wavelet confidence levels are higher than the local 95\% wavelet confidence level. It should be noted that a similar type of findings is reported in previous works for coronal signals (\citealt{Auch2016, Thr2017}). The estimated values of various parameters (e.g., A, s, and $\chi^2$) are also provided in each figure. The low $\chi^2$ values justify the reliability of the fitting model. Panel a and c are dedicated for AIA 1700 while panel b and d are for IRIS 1400 TS. Through the described methodology, we have estimated wavelet power maps for AIA 1700~{\AA} and IRIS 1400~{\AA} and associated 95\% confidence levels (i.e., local $\&$ global wavelet, global Fourier) in the plage region. Figure~\ref{fig:wavelet} shows the normalized wavelet power map with the 95\% wavelet global confidence level as white lines (panel b and d). It shows that power is inherited over a broad range of periods within the photosphere and TR.\\
Further, the wavelet coherence analysis is performed using AIA 1700~{\AA} and IRIS 1400~{\AA}, which is important to understand coherent and incoherent oscillations at two different heights. In the very first step, the cross spectrum (cross wavelet power) of two TS (i.e., photospheric and TR) is determined by multiplying the wavelet of one TS with the complex conjugate of the wavelet of another TS. The cross-spectrum highlights the power areas in the time-frequency range, which has common power in both TS. However, it should be noted that the cross spectrum does not necessarily possess very high power as it is visible in the individual TS.
  \begin{figure*}
   \centering
   \includegraphics[trim = 1.5cm 2.0cm 2.0cm 1.5cm,scale=1.1]{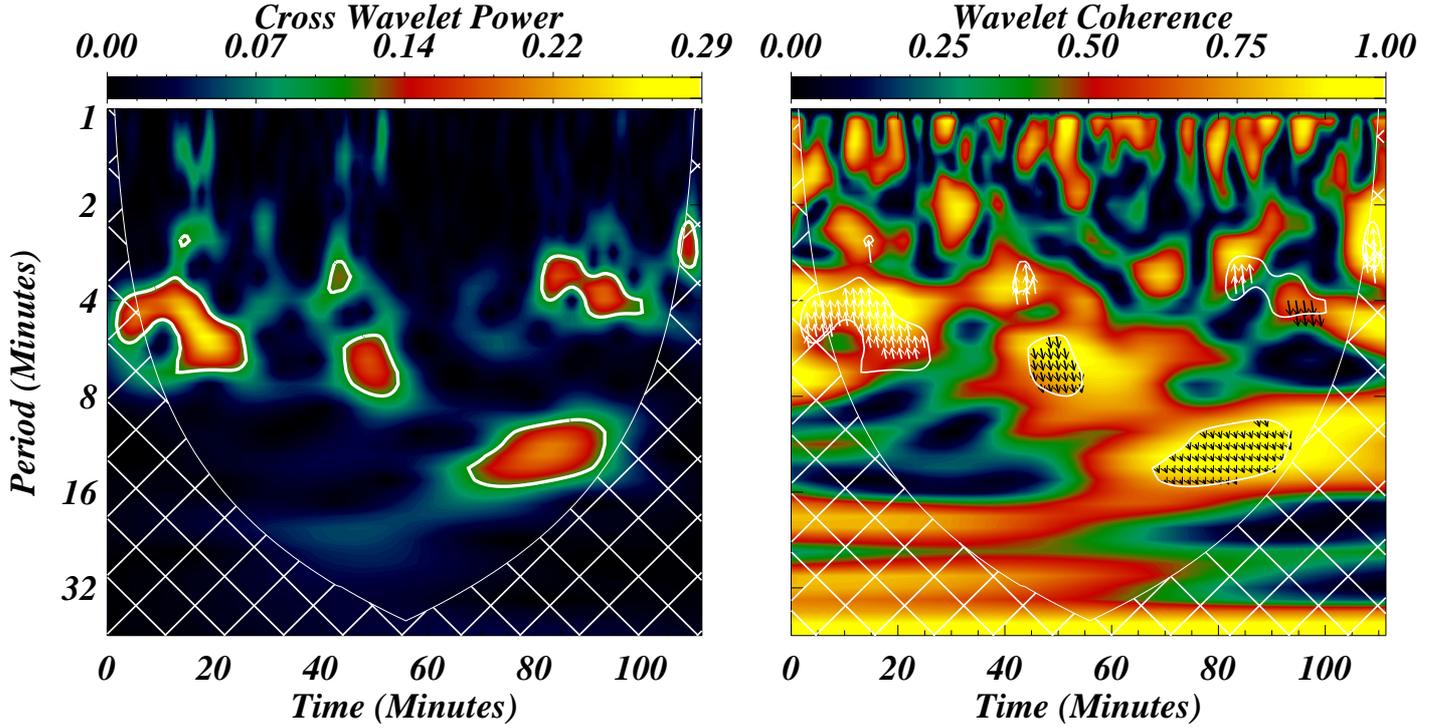}
   \caption{Left: The normalized cross wavelet power between AIA 1700~{\AA} and IRIS/SJI 1400~{\AA}. The normalized cross power is distributed in the similar period range as in the case of wavelet power of photosphere and TR (cf., Fig.~\ref{fig:wavelet}). Right: Wavelet coherence is displayed along with the over plotted phase difference angle (black and purple arrows) within the high coherence regime (i.e., coherence above 0.6). Negative phase difference (black arrows) represent downward propagating waves, while positive phase difference (purple arrows) indicates upwardly propagating waves. It is clearly visible that below 7.0 minutes the waves are propagating upward into the TR from the photosphere. The white contours outline the 95\% confidence levels while white cross hatched area outlines the COI.}
    \label{fig:coherence}%
    \end{figure*}
The left-panel of Fig.~\ref{fig:coherence} shows the normalized cross spectrum, which is drawn using the intensity TS of photosphere and TR. We have utilized the cross FFT power and time-averaged cross wavelet power to estimate the 95\% global confidence levels for this cross wavelet power which is further normalized by noise arrays (as described in subsection 3.1). We have adopted a similar methodology as we applied for individual TS. Figure~\ref{fig:cross_noise} shows two different samples for estimation of the 95\% local and global confidence levels. The white line outlines the normalized cross wavelet power, which is above the 95\% confidence level. The normalized cross wavelet power shows that a significant common power lies in the very wide range of periods, which is a signature of possible relation between the photosphere and TR. However, it should be noted that common cross power alone is not sufficient in order to draw definite conclusions about wave propagation.\\
Therefore, we further study the wavelet coherence, which is an important parameter to investigate the interaction between two TS. The cross wavelet power is normalized by the multiplication of the power of both series, which is basically the wavelet coherence (\cite{Torr1998}). The coherence can vary from zero to one. A value of zero value represents complete incoherence and a value of one the perfect coherence between the two TS. The cross-spectrum of the TS is necessary to find out significant common power in the time and frequency domains, however, the wavelet coherence is also needed to find the relationship between the two heights in the solar atmosphere. Therefore, the wavelet coherence is an important parameter to signify the correlations of oscillations between two heights. Right-panel of Figure~\ref{fig:coherence} shows the corresponding coherence map. Similar to normalized cross power map, coherence map also outlines 95\% confidence levels by white lines. In addition, hatched white area outlines the COI area on cross power and coherence maps.\\
Finally, we have estimated the phase lag (i.e., difference of phase angles at two different heights) in the time and period domains. The cross wavelet gives the complex array in the time-period domain, which can be converted into the phase angle using real and imaginary parts of the complex numbers (\cite{Bloomfield2004, Jafer2017}). The phase angle reflects the phase lag between the two intensity TS originating from two different heights. We have put some specific conditions to select the valid phase lags for the greater reliability of the results, e.g., (a) we have used only those regions which have significant cross power (i.e., cross power lies within the regime of 95\% global confidence level), (b) the coherence value should be greater than 0.6 in these significant cross power areas, (c) we have excluded the COI area to avoid spurious edge effects. Therefore, we have extracted the valid phase lags using the above described conditions in the time-period domain. We have used arrows to show these valid phase lags on the time-period domain. These valid phase lags are over plotted on the coherence maps using white (positive phase lag) and black arrows (negative phase lags). The positive phase lags indicate the upward propagation, while the negative phase lags represent the downward propagation of the waves between two heights. The phase lag is positive in the short period  regime (i.e, less than 6.0 minutes), which shows the propagation of the waves from photosphere to TR (right-panel; Fig.~\ref{fig:coherence}). However, negative phase lag is evident within the regime of high period (i.e., beyond 8.0 minutes). This observational findings indicates that the waves, within the period range of 2.0 to 6.0 minutes propagate upward from the photosphere and can reach up to the TR.\\
  \begin{figure*}
   \centering
   \includegraphics[trim = 1.5cm 0.0cm 2.0cm 0.0cm,scale=1.0]{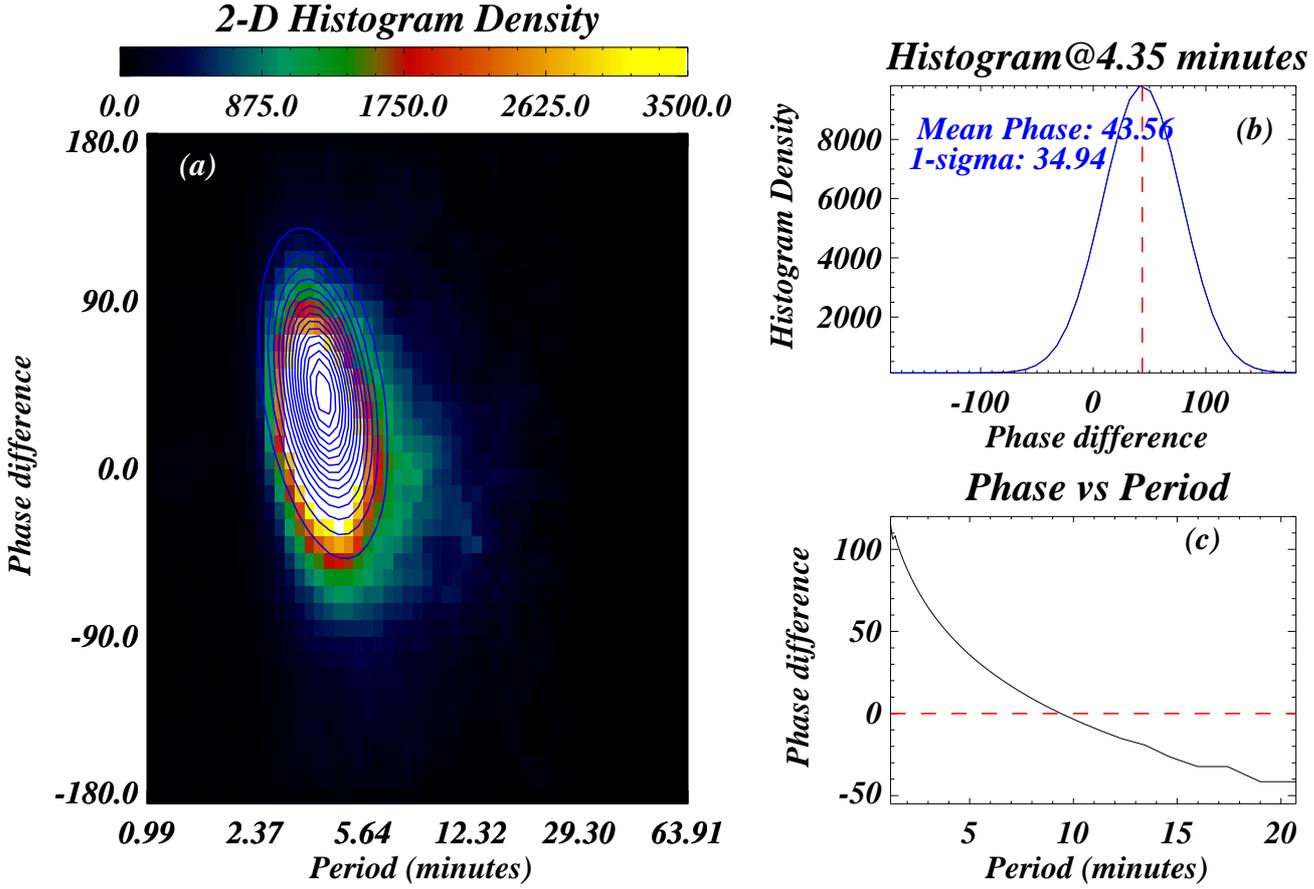}
   \caption {(a) 2-D histogram of $\Delta\,\phi$, extracted from all plage locations. The white contours show a fitted 2-D Gaussian on the 2-D $\Delta$ $\phi$ histogram. The histogram reveals that shorter periods have positive phase difference while longer periods move into the negative phase difference regime. With the help of a 2-D Gaussian, we have estimated the peak phase difference for each period. (b) The distribution of phase difference at the period of 4.35 minutes. (c) Phase difference as a function of period. A vertical dashed line in panel b marks the position of peak phase difference. A dashed horizontal line on panel c marks zero phase difference.}
    \label{fig:histogram}%
    \end{figure*}
We applied the same procedure to all the plage locations to extract valid phase lags. After extracting the phase lags we produce the 2-D histogram (i.e., phase lag vs period) using all plage locations. Panel (a) of Fig.~\ref{fig:histogram} shows a 2-D histogram of phase lags. The histogram reveals the presence of significant phase lag within the period range from 2.0 to 9.0 minutes. Periods less than 7.0 minutes are dominated by the positive phase angles while for periods longer than 7.0 minutes the phase angles are distributed both in the positive and negative regimes. The over plotted blue contours are fitted 2-D Gaussian which successfully characterize the behavior of the observed phase lag distribution. With the help of 2-D Gaussian fitting, we have estimated the peak phase angle at every period. Panel (b) in  Fig.~\ref{fig:histogram} shows the distribution of phase angles at a period of 4.35 minutes. At this period, the peak of the distribution lies around 43.56$^{o}$ with the 1-$\sigma$ spread of 39.94$^{o}$ (see; panel (b)). Finally, we have produced the behavior of mean phase lag with periods that is shown in panel (c) of Fig.~\ref{fig:histogram}. The phase lag is decreasing with increasing period. It is clearly visible that zero phase angle occurs around 9.0 minutes (cf, panel c as well as panel a), which predicts that waves with periods less than 9.0 minutes successfully propagate from the photosphere into the TR.
\subsection{Evidence of the Propagation of Magneto-acoustic Waves}
In this work we aim to investigate the propagation of waves within VMPL through phase difference ($\Delta\phi$). The phase difference between the photospheric (AIA 1700) and TR (IRIS/SJI 1400~{\AA}) reveals the presence of positive phase lag within the period-regime from 2.0 to 9.0 minutes (cf., Fig.~\ref{fig:histogram}). $\Delta\phi$ is decreasing with the period. Such observational finding suggests that most of the photospheric power leaks into the TR through the wave propagation in this period regime. We have performed some further investigations to understand the nature of these waves. Using $\Delta\phi$ and the corresponding frequency and height difference between two atmospheric layers, we can estimate the wave travel time and propagation speed. For instance, the $\Delta \phi$ is around 63.20$^{o}\pm$34.90$^{o}$ at the frequency of 5.4 mHz (or around 3.0 minutes; cf., Fig.~\ref{fig:histogram}).
\begin{equation}
\label{eq:time}
{\tau} = {{\Delta \phi}\over {2\pi f}}, 
\end{equation}
Using Eq.~\ref{eq:time} and upper value of $\Delta\phi$ (i.e., $\mu$+2.0$\sigma$), the wave travel time ($\tau$) is around 68.34 s for the frequency of 5.4 mHz. The Si~{\sc iv} 1393.77~{\AA} spectral line typically forms at a geometrical height of 2-3 Mm above the photosphere. Lets assume a height difference of 2 Mm (lower bound value) between the photosphere (AIA 1700~{\AA}) and TR (IRIS/SJI 1400~{\AA}). So, the wave travel time and distance (height difference between photosphere and TR) yields a wave propagation speed of $\sim$29 km s$^{-1}$. 
It is believed that waves travel with slightly lower speed (i.e., 15-20 km/s) within the lower solar atmosphere (B. De Pontieu; private communication) and our estimated propagation speed is higher than expected. We have used VMPL (very bright and high-magnetic field areas) that can have significantly higher temperatures compared to the quiet-Sun. We do not have the temperature measurement in present baseline, although surely, qualitatively we can assume that plage should have higher temperatures compared to the quiet-Sun. These higher temperature can lead to a higher sound speed. So, we may infer that the speed of the propagating waves is close to the sound speed within the plage region. Our observational findings indicate that these are slow magneto acoustic waves (SMAW). \cite{DePon2003,DePon2005} have also reported the propagation of SMAW into the corona from the photosphere in plage regions. Nonlinear effects contribute to complex wave dynamics \citep{Hegg2011,Sko2016} while the occurrence of shocks produce strong gradients in  various physical parameter (i.e., intensity or Doppler velocity) over a short period of time (\cite{Tian2014}). Variations of the physical parameters (i.e., intensity or Doppler velocity) should reveal the saw-tooth pattern as reported by \cite{Tian2014}. Panel (b) of figure~\ref{fig:SJI_SI} shows the Si~{\sc iv} 1393.77~{\AA} intensity images (y-t image).

A large number of samples (i.e., 5927 TS within the plage) are utilized in the present analysis. We have found that a large number of locations show good correlation between the photosphere and the TR over a significant time interval. That is the reason we have used all the locations to create the final 2-D histogram (cf., Fig~\ref{fig:histogram}). Previous reports state that only a small fraction of the locations show a nice correlation between two atmospheric layers. For example, \cite{DePon2003} have reported that only 4\% locations (6 out of the total 148) show a good correlation that indicate the successful propagation of the SMAW between the photosphere and TR. In the present observation, we have found that almost all locations have correlation between photosphere and TR. In previous work, with the help of FFT, \cite{Centeno2009} have utilized all the locations to draw the phase difference diagram (frequency vs phase) between photosphere and chromosphere. We have also used all the locations to draw the phase vs. frequency histogram. In this work, we use IRIS high resolution observations and analysis techniques (wavelet, cross spectrum, coherence, and a generic noise model), which lead these results with even more accuracy. Higher occurrence of propagation of SMAW within the plage region are inferred from IRIS observations for first time. We believe that this is important for the energy transport and thereby possible heating of the atmosphere overlying the plage region.
\section{Discussion $\&$ Conclusions}
\cite{Centeno2006,Centeno2009} investigated wave propagation in different features of the solar atmosphere (e.g., umbra, pore and faculae) through $\Delta$ $\phi$ using FFT. They reported a successful propagation of 3-minute oscillations into the chromosphere for each of these solar features. In the case of faculae, they found that low frequency waves (5 minute) can also reach the chromosphere.  The variation of cut-off frequency was also investigated using frequency vs $\Delta\,\phi$ diagrams (\citealt{Centeno2006, Centeno2009}). Their findings suggest that waves with longer periods can also propagate into the TR in a plage environment. \cite{Centeno2006,Centeno2009} used only vertical magnetic field and successfully managed to show the propagation of 5-minute oscillations in faculae. \cite{DePon2004} reported that the leakage of p-modes  into the upper atmosphere is only possible along inclined magnetic fields while \cite{Centeno2009} reported the propagation of 5-minute waves in faculae for non-inclined (almost vertical) magnetic field. \cite{Hegg2011} used numerical simulations to study wave propagation. Waves with a 5 minute period dominate in strong and inclined flux-tubes whereas 3 minutes dominates in weak or vertical magnetic flux tubes. Similar to \cite{DePon2004}, \cite{Hegg2011} have also found that flux tube inclination is important for long-period wave propagation. Further, \cite{DePon2003,DePon2005} have reported that longer period waves associated with the photospheric 5 minute oscillations can reach into the TR within the plage region. 
    
  In the present work we have estimated 2-D Fourier power maps from plage observations (i.e., plage+surrounding area) at different frequency ranges (i.e., high to low-frequencies). The Fourier power maps show how the wave power is distributed in the plage $\&$ surrounding area and show that most of the wave dynamics are inherited in the intermediate frequency range. We found very low power at high frequencies without any significant difference in the plage and surrounding region. Similarly, low frequencies do not show the difference between the plage and surrounding region but have high power in comparison to the power inherited in the high frequency range. Hence, on-average, the Fourier power behavior suggests that the dynamics of the plage region is inherited within the regime of intermediate frequencies. Then, we applied for the first time a generic noise model to estimate the confidence levels using wavelet analysis on photospheric and TR TS. The noise model has been previously applied only for coronal TS \citep{Auch2016,Thr2017}. Previous scientific works have utilized either a white or red noise model to calculate confidence levels, which can give erroneous confidence levels. In addition, \cite{Auch2016} have reported that smoothing can include spurious periods, therefore, we did not apply any type of smoothing to our  TS. The generic noise model and original TS  have significantly improved the accuracy of the results. Our analysis has revealed that waves within a certain period range (i.e., 2.0 to 9.0 minutes) are propagating into the TR from the solar photosphere. We have found that $\Delta\,\phi$ decreases with period, which qualitatively matches with the previous reported results (e.g., \citealt{Centeno2006,Centeno2009}). 

We have estimated the wave travel time and propagation speed using $\Delta\,\phi$, frequency and height difference (between photosphere and TR), e.g. 68.34 seconds wave travel time at a frequency of 5.4 mHz, which gives the propagation speed of about 29.26 km s$^{-1}$. This propagation speed is close to the sound speed in the TR, which justifies our conclusion that these are SMAWs. In addition, a significant correlation/propagation (in terms of the locations) between the photosphere and TR is revealed in the present analysis for the first time using IRIS and AIA observations. Previous works report very small fraction of a good  correlation between photosphere and TR \citep{DePon2003}. The occurrence frequency of correlation/propagation of SMAW is very high in the present study. It should be noted that we have only used the almost vertical magnetic field plage locations and find that a broad range of waves (i.e., 2 to 9 minutes) successfully reach into the TR from photosphere. Finally, our results suggest that the propagation of 5 min oscillations may not depend on the magnetic field inclination, which supports the findings of \cite{Centeno2006,Centeno2009}. We have investigated the nonlinear character of waves (i.e., shocks) and did not find any such signature.\\   

We investigated which emission (i.e., photospheric {--} continuum part of SJI 1400~{\AA} filter or Si~{\sc iv} lines in SJI~1400~{\AA}) dominates in the SJI~1400~{\AA}. On the basis of a light curve comparison and phase analysis, we found that the SJI~1400~{\AA} emission is dominated by the Si~{\sc iv} lines (see appendix~A for more details). Similarly, \cite{Sko2016} have reported that bright grains emission originates from the TR (i.e., Si~{\sc iv} lines rather than the photosphere (i.e., continuum part of SJI 1400 filter). However, using the same observations, \cite{Marty2015} suggests that the emission of bright grains originates from the photosphere.
\begin{acknowledgements}
\textbf{We sincerely thank Dr. Clara Froment for her constructive comments that improved the paper significantly.} We would like to thank Dr. Bart De Pontieu (LMSAL) for helpful comments on a previous version of the paper. IRIS is a NASA small explorer mission developed and operated by LMSAL with mission operations executed at NASA Ames Research center and major contributions to downlink communications funded by ESA and the Norwegian Space Centre. S.K.T. gratefully acknowledges support by the NASA contract NNG09FA40C (IRIS). AKS and MM acknowledge the support of the UGC-UKIERI grant to support the present research.
\end{acknowledgements}

\bibliographystyle{aa}
\bibliography{listb}
\begin{appendix}\label{append_main}
\section{Consistency between IRIS/SJI 1400~{\AA} and Si~{\sc iv} 1393.77~{\AA}}
As we have derived well correlated oscillations in near photospheric (AIA 1700~{\AA}) and TR (IRIS/SJI 1400~{\AA} emission), we need to evaluate is the IRIS/SJI 1400~{\AA} bandpass is indeed dominated by the TR emission. The IRIS/SJI 1400~{\AA} filter is a 55~{\AA} broad filter, which contains two strong Si~{\sc iv} lines. However, a large portion of this filter is dominated by the photospheric continuum.  The Si~{\sc iv} 1393.77~{\AA} spectral line, which is the strong line within this filter, is also included in for this observation. With the help of this spectral line, we can check the dominance of emissions within this filter.\\
  \begin{figure*}
   \centering
   \includegraphics[trim = 1.0cm 0.0cm 2.0cm 0.0cm,scale=1.1]{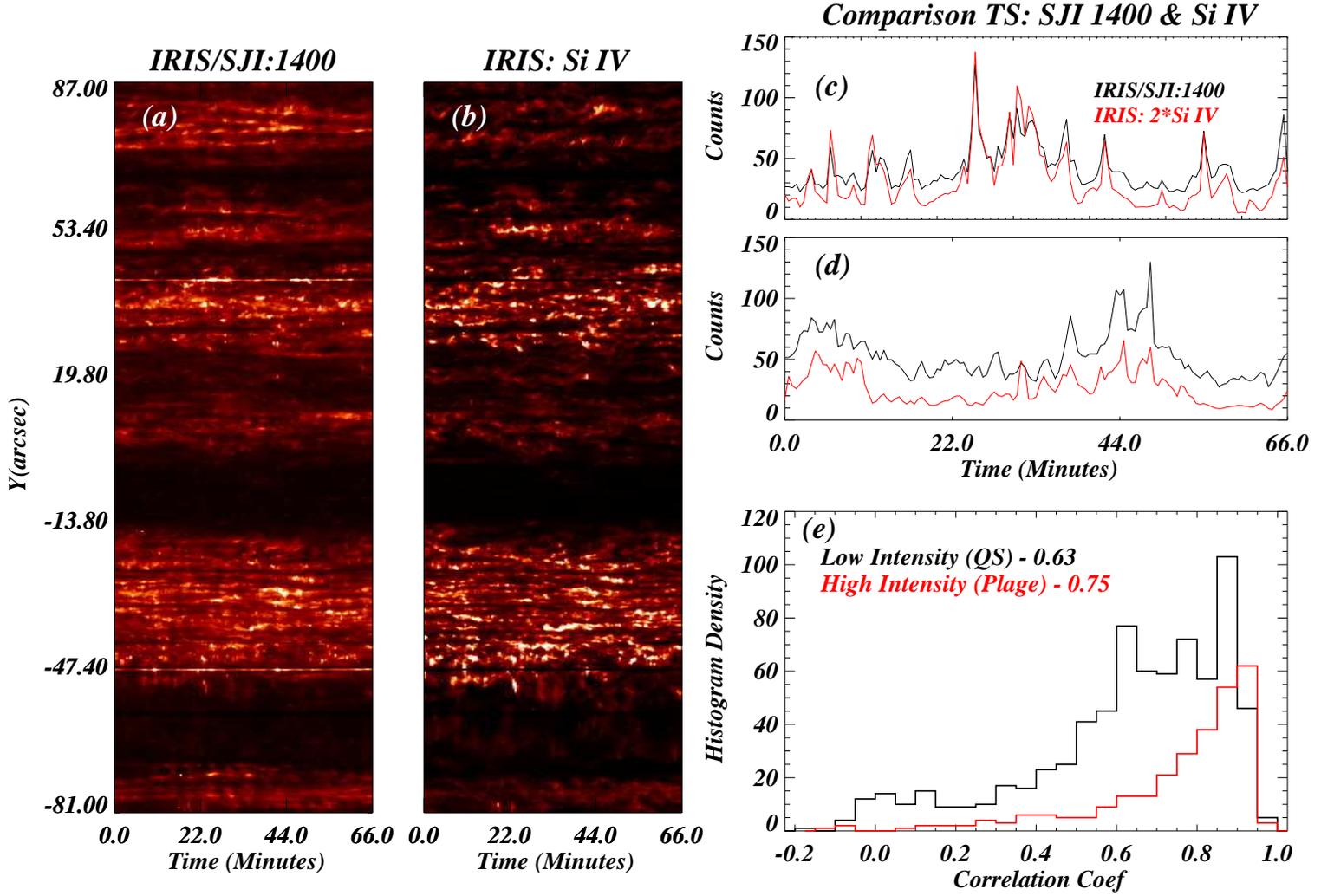}
   \caption{Panels (a) and (b) show the y-t intensity images from IRIS/SJI 1400~{\AA} and Si~{\sc iv} 1393.77~{\AA}. These images are very similar. We have compared IRIS/SJI 1400~{\AA} TS (black line) with Si~{\sc iv} 1393.77~{\AA} (2 times Si~{\sc iv}) from two different locations (panels (c), and (d)) to understand the similarities/differences in the nature of both TS.  Panel (e) shows histograms of the correlation coefficients for QS (black) and plage (red).  The plage histogram is sharply peaked towards high values.}
    \label{fig:SJI_SI}%
    \end{figure*}
First, we estimate the intensities from the Si~{\sc iv} 1393.77~{\AA} line using Gaussian fitting. By selecting the slit position in SJ images, we have produced y (along the slit)-t(time) intensity images from IRIS/SJI 1400~{\AA} data cube. This allows for a comparison between the y-t intensity images of SJI 1400~{\AA} and Si~{\sc iv} 1393.77~{\AA}. The intensity image from SJI 1400~{\AA}) is very similar to the intensity image from Si~{\sc iv} 1393.77~{\AA} (cf., panel a and b of Fig.~\ref{fig:SJI_SI}). We have also examined TS from two different locations within the plage (cf., panels c and d; Fig~\ref{fig:SJI_SI}). This shows that the SJI 1400~{\AA} emission exactly follows the TR emission (emissions from Si~{\sc iv} 1393.77~{\AA}). Using intensity threshold, we divide the regions into plage and surrounding quiet-Sun (QS). We then look for correlations between the IRIS/SJI 1400~{\AA} and Si~{\sc iv} 1393.77~{\AA} TS for each location. Panel c of Fig.~\ref{fig:SJI_SI} shows the statistical distribution of correlation coefficients for plage (red histogram) and QS (black histogram). The histograms reveal that the QS coefficients show a very large spread while the plage histogram is  sharply peaked at a high value (around 0.7). It is also evident that the mean coefficient is high for plage region (0.75) compared to QS (0.6). These observational findings support that SJI 1400~{\AA} and Si~{\sc iv} 1393.77~{\AA} are highly correlated with each other in the plage regions.\\
  \begin{figure*}
   \centering
   \includegraphics[trim = 1.0cm 0.0cm 1.0cm 0.0cm,scale=1.1]{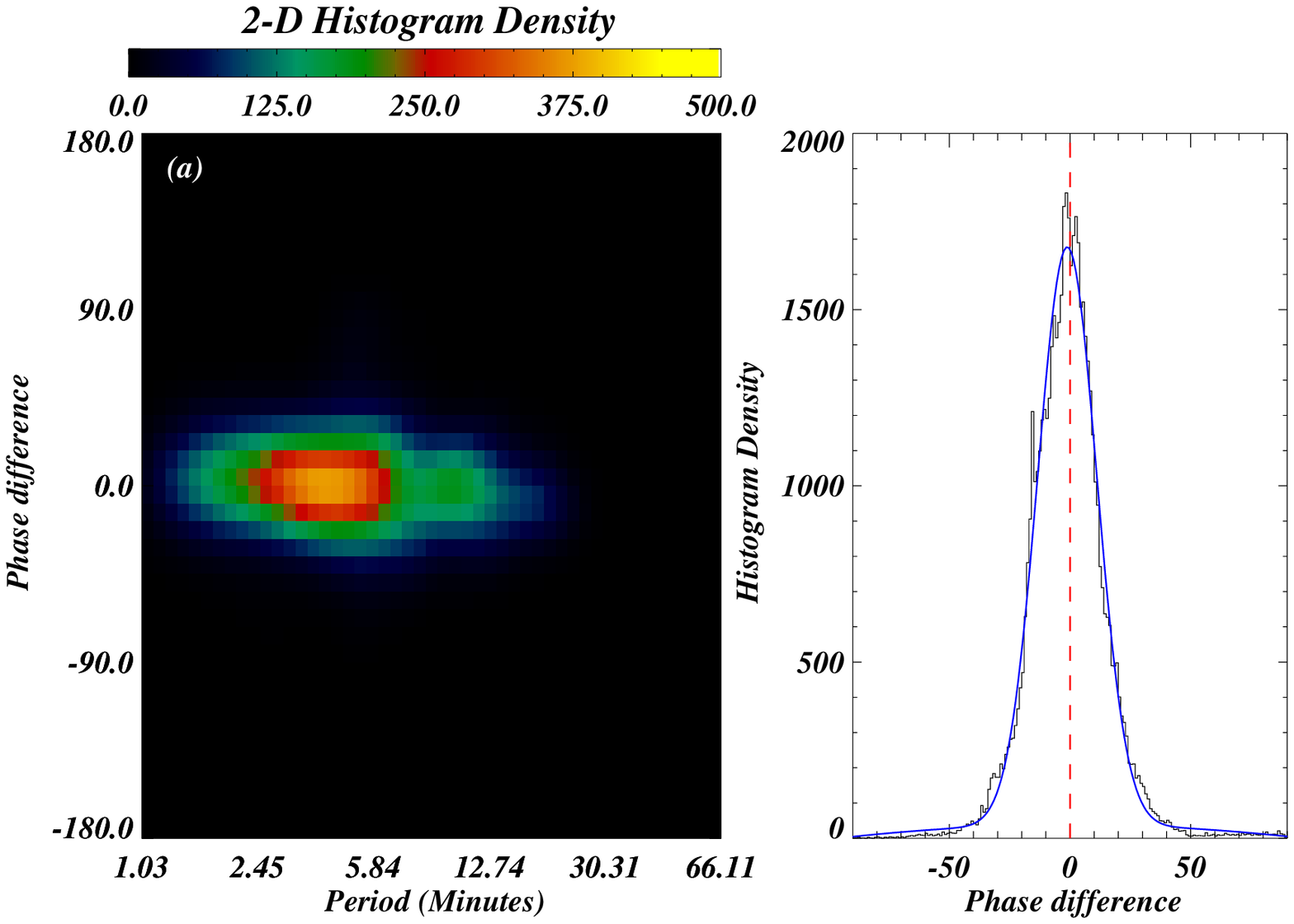}
   \caption{Panel (a) shows a 2-D histogram of phase differences which is distributed within $\pm$ 20. panel (b) shows the distribution of phase difference (black line), which are estimated using IRIS/SJI 1400~{\AA} and Si~{\sc iv} 1393.77~{\AA}, along with its Gaussian fitting (blue line).The statistical distribution peaks around a 0 phase difference with a FWHM of 20. A vertical dashed line marks the location of zero phase difference. }
    \label{fig:SJI_SI_PHASE}%
    \end{figure*}
In the further step, we have investigated the phase difference between IRIS/SJI 1400~{\AA} and Si~{\sc iv} 1393.77~{\AA}. The Si~{\sc iv} 1393.77~{\AA} line forms in the TR. If IRIS/SJI 1400~{\AA} is dominated by the emission originating from the TR, then the phase lag between SJI 1400~{\AA} and Si~{\sc iv} 1393.77~{\AA} should be around zero phase.  Applying the same methodology (as previously described), we have estimated the phase lag between SJI 1400~{\AA} and Si~{\sc iv} 1393.77~{\AA} and this is shown in Fig.~\ref{fig:SJI_SI_PHASE}. The 2-D histogram (a) shows that almost all phase angles are accumulated within $\pm$15 at each period. In addition, the phase lags have no dependency on the periods as it should be expected theoretically and we found previously. In (b) we also show the statistical distribution of the estimated phase differences. The black line is a histogram of phase lags, while the over plotted blue line is the Gaussian fit of the histogram. The distribution shows that the peak lies around zero phase difference (i.e., -1.16). Moreover, the Gaussian fitting shows that the Full Width at Half-Maximum (FWHM) is around 20. On the basis of these observational findings, we can say that IRIS/SJI 1400~{\AA} captures the emissions from TR within the plage areas.
\end{appendix}
	
\end{document}